# Probing the polar-nonpolar oxide interfaces using resonant x-ray standing wave techniques

Cheng-Tai Kuo,[1,a] Shih-Chieh Lin,[2] and Yi-De Chuang [3]

[1]Stanford Synchrotron Radiation Lightsource, SLAC National Accelerator Laboratory, Menlo Park, California 94025, USA
[2]Department of Physics, University of California Davis, Davis, California 95616, USA
[3]Advanced Light Source, Lawrence Berkeley National Laboratory, Berkeley, California 94720, USA

[a] Electronic mail: ctkuo@slac.stanford.edu

Transition metal (TM) oxide heterostructure superlattices have attracted great attention in research communities because of their emergent interfacial phenomena that do not exist in the bulk form. In order to understand the mechanisms that cause these phenomena, it is important to use depth-resolved spectroscopies to study the electronic structure across the buried oxide interfaces. In this review, we focus on the recent applications of standing wave (SW) photoemission (SW-XPS) and resonant inelastic x-ray scattering (SW-RIXS) spectroscopy to study the depth profiles of electronic structure or carriers around the polar-nonpolar oxide interfaces. Using the incident photon energies near the TM x-ray absorption resonance, the created SW excitation can enhance the spectral response and certain electronic transitions, providing important insight into the interfacial electronic structure in the energy and real space regimes. Following the background introductions, we describe two SW experiments and demonstrate that the combination of SW-XPS and SW-RIXS has the potential to obtain the depth distribution of electronic/orbital states around the buried interfaces with Angstrom precision.

# I. INTRODUCTION

Transition metal oxide (TMO) perovskite thin films exhibit a broad spectrum of physical properties.[1,2,3] In their generic formula $ABO_3$, the high solubility of different A-site and B-site metal cations offers an abundant phase space for the compositions and the resulting electronic/magnetic structures. Thanks to the advanced thin film growth techniques such as pulsed laser deposition (PLD) and molecule beam epitaxy (MBE), many new artificial heterostructure superlattices (SLs) with emergent physical properties that do not exist in the bulk form were discovered over the past decades.[4-8]

Among these emergent properties in TMO heterostructures, what has attracted the most attention is the novel conductivity at certain interfaces of polar and nonpolar perovskite insulators. One renowned example is the $LaAlO_3/SrTiO_3$ (LAO/STO) heterostructure because its interfacial two-dimensional electron system (2DES) possesses superior electronic and magnetic properties.[9-12] In these polar-nonpolar heterostructures, some may exhibit interfacial conductivity (*e.g.* $GdTiO_3/SrTiO_3$[13], $NdTiO_3/SrTiO_3$[14,15], etc.) while others do not (*e.g.* $LaCrO_3/SrTiO_3$ (LCO/STO)[16,17]). For $LaFeO_3/SrTiO_3$ (LFO/STO), the insulating interfaces[18,19] and 2DES[20] were reported, depending on the conditions of interfaces (e.g. oxygen stoichiometry). The distinction may be related to the contrasting nature of the electronic structure around the interfaces. For example, Chambers *et al.* found that the LCO/STO heterojunctions were not conductive, and there was evidence of a built-in potential within the LCO layer from the interfacial polarization discontinuities that suppresses the conductivity.[16] It is worth noting that Dr. Scott A. Chambers' group developed a powerful technique, the *in-situ* x-ray photoelectron spectroscopy (XPS), to study the emergent phenomena in various polar-nonpolar oxide heterostructures.[16-19,21-25]

This depth-resolved spectroscopy allows probing the interfacial composition and electronic structure with positional precision down to one unit cell (uc) for each constituent layer. Another technique is the standing-wave-excited (SW) XPS, which involves generating an SW within the sample to modulate the emission of photoelectrons from a selected depth. With this feature, SW-XPS is a promising technique because it can highlight the spectral contribution from individual interfaces by judicially tuning the intensity of the SW electric field at the prescribed depth in the sample.

In this review article, we will discuss how the novel interfacial phenomena at the polar-nonpolar oxide interfaces can be investigated using the x-ray SW spectroscopic techniques, such as SW-XPS and SW- resonant inelastic x-ray scattering (RIXS). We first introduce the basic concept of SW techniques and explain how XPS/RIXS can attain depth sensitivity when combined with the SW excitations. We then discuss how to explore and design a superlattice sample that is suitable for SW experiments, in particular, using the incident x-rays with energies tuned to the TM x-ray absorption (XAS) edge to enhance the SW effects for specific transitions in the XPS/RIXS processes. In the following sections, we review the SW-XPS and SW-RIXS experiments on two interesting heterostructures: The LAO/STO and LCO/STO. In the last section, we summarize these findings and offer a future perspective on the SW-XPS/RIXS techniques, as well as some promising polar-nonpolar systems that can benefit from these techniques.

## II. X-ray standing wave spectroscopies

SW-XPS is a powerful, versatile, and non-destructive technique for probing the element-specific electronic, magnetic, and structural properties of buried layers and interfaces with sub-nanometer depth resolution.[26-28] C. S. Fadley,[26-28] J. Zegenhagen,[29-34] and M. Bedzyk[35-39] have made significant contributions in the developments of x-ray SW techniques. The early SW experiments were carried out on bulk crystals using hard x-rays that match the length scale of the natural crystallographic atomic planes to generate the SW excitation.[40-42] In the 2000s, Yang and Fadley *et al.* developed a new type of depth-resolved photoemission technique where the SW was generated by a synthetic superlattice with periodicity on the order of several nanometers. This increased length scale extends the SW technique into the soft x-ray regime.[43-45] To assist the interpretation of SW experimental results, a program called Yang X-ray Optics (YXRO) was developed and made available for the public to simulate the x-ray optical effects in various SW spectroscopies.[46,47] In 2010, Gray *et al.* first applied the SW-XPS to an TMO SL (LaSr$_{0.33}$Mn$_{0.66}$O$_3$ (LSMO)/STO) and investigated the interfacial magnetic phenomena.[48] Recently, SW-XPS experiments were performed on various oxide SLs[49-54] and the freestanding quantum membranes on a multilayer mirror.[55] Besides SW-XPS, SW-RIXS is a newly developed technique that combines the strength of RIXS in probing the orbital and magnetic excitations and that of the SW excitation in depth resolution. Kuo *et al.* reported the first demonstration of SW-RIXS characterization of a La$_{1.85}$Sr$_{0.15}$CuO$_4$ (LSCO)/LSMO SL.[56] They observed the SW effects on the *dd* excitations and showed the different depth profiles for the *dd* and magnetic excitations in these LSCO/LSMO multilayers. In 2020, Lin and Kuo *et al.* also applied the SW-RIXS to study the 2DES in an LAO/STO SL to elucidate its nature.[54] With these pioneer works, the SW spectroscopies (SW-XPS/RIXS) now focus on controlling the interplay between the SW electric field and the interfaces of interests by setting the incident photon energy near the TM XAS resonance, which is the main focus of this review article.[52, 54, 56]

In these SW spectroscopies, the SW is created inside the sample by a Bragg reflection from a multilayer, and the depth-dependent SW electric field modulates the emission of photons or photoelectrons.[47] In Figure 1(a), we use an LSCO/LSMO SL to illustrate the relevant parameters for SW spectroscopic experiments. The LSCO/LSMO SL sample, denoted as (LSCO$_2$/LSMO$_7$)$_{20}$, consists of 20 repetitions of 2 uc LSCO and 7 uc LSMO with periodicity ($d_{ML}$) of ~ 53.4 Angstroms. When varying the incidence angle $\sin\theta_{inc}$, the SW is scanned through the first-order Bragg condition $\lambda_x = 2d_{ML}\sin\theta_B$. Here, $\lambda_x$ is the wavelength of the incident x-rays and $\theta_B$ is the incident angle for the first-order Bragg reflection. The SW electric field intensity varies sinusoidally with sample depth and its periodicity is very close to $d_{ML}$. This can be seen in the intensity map of the calculated SW electric field as a function of sample depth and $\sin\theta_{inc}$ (Figure 1(b)). In this figure, we see the SW moves vertically as the incidence angle is varied across the Bragg angle (~ 8°). For this specific SL, the maximum of SW electric field is located at the first LSCO/LSMO interface.

When spanning the whole Bragg peak, the SW shifts vertically by half a cycle and the modulating electric field enhances or reduces the emission of photons or photoelectrons at different depths. The intensity of emitted photoelectrons or photons as a function of incidence angle, which is what we call a rocking curve (RC), is described as

$$I(\theta_{inc}) \propto 1 + R(\theta_{inc}) + 2\sqrt{R(\theta_{inc})}f\cos\left[\varphi(\theta_{inc}) - 2\pi(z/\lambda_{SW})\right]. \qquad (1)$$

Here, $R(\theta_{inc})$ is the reflectivity of SL, $f$ is the fraction of atoms in the coherent positions, $\varphi(\theta_{inc})$ is the phase difference between the incident and scattered waves, and $z$ is the depth. The amplitude modulation of an SW excitation in $I(\theta_{inc})$ is proportional to the square root of the reflectivity $R(\theta_{inc})$, which means that an oxide SL with higher reflectivity exhibits a stronger SW effect in the intensity modulation.

To determine whether the desired oxide heterostructure is suitable for an SW characterization or not, as a first step, one needs to know its reflectivity. The reflectivity of an SL at a given incident photon energy depends on the sample structure, the indices of refraction of the constituent materials, the sharpness of the interfaces, etc.[47] With these variables that can affect the SW effects, it is instructive to examine their relative contributions. The reflectivity of an SL can be calculated from the Center for X-ray Optics (CXRO) website.[57] In Figure 2(a), we show the comparison of the reflectivity of an LCO/STO SL with various sample geometries at 900 eV incident photon energy. These samples all have 10 repetitions, but within each repetition, they have different numbers of LCO and STO unit cells. From this figure, we see that the reflectivity peak positions of $(LCO_5/STO_{10})_{10}$ and $(LCO_{10}/STO_5)_{10}$ are very close to each other because they have very similar periodicities ($d_{ML}$). It appears that the $(LCO_5/STO_{10})_{10}$ SL has the strongest reflectivity around 4%, which gives an SW modulation of ~ 40%. Figure 2(b) shows the reflectivity of an $(LCO_5/STO_{10})_{10}$ SL at different incident photon energies. When increasing the photon energy from 700 to 1100 eV, the peak position shifts to a lower angle and the peak width becomes narrower. In addition, the reflectivity at 1100 eV is the highest. To have a successful SW experiment, it is prudent to choose a good combination of sample geometry and incident photon energy. In principle, one can use a simulated reflectivity of ~ 0.01 (1%) for an SL as the criterion because this SL is expected to exhibit ~ 20% SW modulation in the emission of photoelectrons or photons. But in practice, the surface/interface roughness and interdiffusion often reduce the reflectivity and SW modulation, thus one needs to start with a higher reflectivity. In this regard, it is interesting to note that the LCO/STO SLs is a particularly ideal system for SW experiments because their reflectivity of 0.2 ~ 0.4 gives relatively high SW modulations.[52]

For certain oxide SL, its reflectivity contrast can be too small for SW experiments if the constituent materials have very similar indices of refraction. One useful trick for enhancing the contrast is choosing the incident photon energy around the absorption edge of one of the constituent materials. In Figure 3 (a), we show the comparison of the reflectivity of an $(LSCO_2/LSMO_7)_{10}$ SL at on-resonant (~932 eV, Cu $L_3$ edge) and off-resonant (900 eV) photon energies. The on-resonant reflectivity of >0.02 is nearly an order of magnitude larger than the ~ $3\times10^{-3}$ off-resonant reflectivity, and this reflectivity enhancement is critical for SW experiments that are otherwise non-feasible if using the non-resonant photon energies. Note that in Figure 3(a), the on-resonant reflectivity comes from the experiment while the off-resonant reflectivity is obtained from the CXRO website.

Although the CXRO website contains the index of refraction ($\beta$ and $\delta$) of elements over a wide photon energy range, the resonance effects are not included in the database. To correctly calculate the reflectivity near the resonant photon energy, the resonant index of refraction ($\beta_{res}$ and $\delta_{res}$) is needed. Figure 3(b) shows the resonant index of refraction of the $(LSCO_2/LSMO_7)_{10}$ SL as a function of photon energy and the values can be derived from the experimental XAS spectra using the Kramers-Kronig analysis. The YXRO program mentioned earlier is capable of producing the resonant reflectivity using resonant index of refraction. This program not only simulates the reflectivity but also produces the SW electrical field map and the photoelectron/photon intensity yields of a defined sample with a large degree of freedom in terms of the materials, sample geometries, stacking sequences, etc. Using this powerful program, we are able to study how the SW electric field interacts with the emissions of photons and photoelectrons via resonant effects, which will be discussed in the next sections.

## III. SW-XPS with incident x-rays near TM XAS resonance

In this section, we describe two SW-XPS experiments that utilized the incident x-rays with energies near the TM XAS edges.[52,54] The first experiment is on an LCO/STO SL.[52] In this experiment, the incident photon energies were chosen around the La $M_5$ edge (Figure 4(a)) where the real (refractive, δ) and the imaginary (absorptive, β) parts of the index of refraction exhibit dramatic variations (Figure 4(b)). Choosing incident photon energies close to this resonance increases the SW sampling depth and enhances the reflectivity, leading to stronger SW effects. In Figures 4 (c) and 4(d), we show the spatial distribution of the SW electric field intensities at photon energies below (829.7 eV) and above (831.5 eV) the La $M_5$ edge, respectively. At 829.7 eV, the maximum of the SW electric field is located at $LCO_{top}/STO_{bottom}$ when the incidence angles are in between 5.5° and 7°. This maximum sweeps down to the middle of the first STO layer above 7° and stays there until the end of the angle scan. We see a similar SW electric field intensity map at 831.5 eV except the maximum is shifted downward by ∼20 Å. The down-shifted SW electric field at this photon energy provides more sensitivity to the second interface STO$_{top}$/LCO$_{bottom}$.

By using the SW excitation and recording the core level and valence band (VB) photoelectron spectra as a function of incident angle from both LCO and STO layers, one can study the interfacial electronic structure in LCO/STO SL. For example, one can visualize the built-in potential drops, which are the potential gradients caused by the polar discontinuities across the polar-nonpolar interfaces in the LCO/STO SL. Figures 4(e) and 4(f) show the relative energy shift of Sr 3$d$ and La 4$d$ core levels as a function of incidence angle at 829.7 eV and 831.5 eV, respectively. The relative energy shift reflects the interplay between the built-in potential gradients and the SW electric field within the SL. We can determine the potential gradients by performing the least-square fitting on the energy shift profiles in Figures 4(e) and 4(f) using a linear form for the built-in potential as a trial input. The determined potential gradients are shown in Figure 4(i), which reveal the +1 eV and −0.8 eV changes in the binding energy along the depth direction in the LCO and STO layers, respectively.

The VB offsets at each interface in Figure 4(i) are further determined by the following analysis on the VB maxima. We show two sets of deconvoluted, matrix-element-weighted density of states (MEDOS) from the LCO and STO layers recorded at the same photon energies in Figures 4(g) and 4(h). In these figures, the curves denoted "simulation" are the modified LCO and STO VB spectra (obtained from bulk LCO and STO reference samples) by considering the potential gradients and the steps at the polar interfaces. The curves denoted "deconvolution" are the derived LCO and STO MEDOSs collected at two different photon energies. By combining the derivation of the slopes of potential gradients within each layer and the magnitude of valence band offsets, the absolute potential values with respect to the VB maxima in both LCO and STO layer are finally determined, annotated as the SW-XPS-derived profile (turquoise curves) in Figure 4(i). We find that this procedure can uniquely determine the potential variations along the depth direction. The analysis suggests that in the 5 uc LCO layer, the VB edge shifts to a higher binding energy by 1 eV while in the 10 uc STO layer, the VB edge shifts to a lower binding energy by 0.8 eV. This finding shows the agreement between the qualitative expectation of the charged interface configuration and the sign of the potential gradients, i.e. the more positively charged interfaces, the higher binding energy for the valence electrons. The detailed theory and discussions of this methodology can be found in Ref. 52.

The second example is the well-known LAO/STO heterostructure.[54] In this experiment, the incident photon energy was around the Ti $L_3$ XAS edge. Note that collecting the photoelectron spectra with incident photon energies around the XAS edge is also called resonant x-ray photoemission spectroscopy (RXPS). Tuning the incident photon energy to Ti $L_3$ edge will enhance the spectral weight of Ti 3$d$ interface states in the RXPS spectra. These interface states, close to the Fermi level, are known to be associated with the in-gap (IG) states from the oxygen vacancies or the quasiparticle (QP) states from the 2DES.[58,59]

Like in the previous example, we can determine the depth profile of Ti 3$d$ interface states using SW-RXPS. Figure 5(a) shows the RXPS VB spectrum with the inset that gives a magnified view around the Fermi level to reveal a near Fermi (NF) peak centered at ~ 0.35 eV. This NF peak contains the contributions from both QP and IG states.[58,59] The experimental NF peak (open circles) and the best-fit (curve) core level photoelectron yields (or the so-called rocking curves, RCs) are shown in Figure 5(b). From this figure, it is evident that both curves exhibit contrasting profiles compared to other core-level RCs like that from La 4$d$ (LAO layer) and Sr 3$d$ (STO layers), suggesting this NF peak has its own depth distribution. Since NF peak is associated with the Ti 3$d$ orbitals, its depth distribution can be related to the Ti 3$d$ interface carriers or the 2DES in the STO layer.[58-60] The determined depth profile at the LAO/STO interface is summarized in Figure 5(c), which shows that in addition to the widely known mobile electrons (e−) in the STO layer (2DES), there is an unexpected high-concentration source of Ti 3$d$ interface carriers in the LAO layer. The mobile 2DES resides within the 3 uc STO layer (~12 Å in length) near the interface, and this length scale is consistent with the recent STEM results of ~10 ± 03 Å.[61] According to the polarity-induced defect mechanism for the 2DES formation, the polarity discontinuity across the LAO/STO interfaces triggers the spontaneous formation of the antisite defects (Ti-on-Al atoms, $Ti_{Al}$) in the LAO layers to alleviate the polarization-induced field across the interfaces.[62] In this scenario, these unexpected Ti 3$d$ interface carriers in the LAO layer can be assigned to the polarity-induced defects. The SW-RXPS results show the strong experimental evidence that supports the polarity-induced defect mechanism.

## IV. SW-RIXS

Ti $L_3$ edge RIXS, which probes Ti 3$d$ orbital transitions ($dd$ excitations) between occupied and unoccupied states, has been proven to be extremely useful for understanding the 2DES in the Ti-based heterostructure.[54,63-67] Combined with SW excitation that gives the depth resolution, we can use this technique to gain further insight into the depth distribution of Ti 3$d$ orbitals ($d_{xz}/d_{yz}$, $d_{x^2-y^2}$, and $d_{z^2}$) and its relationship with the formation of 2DES. Due to the quantum confinement effect,[63,66,68] the half-filled $d_{xy}$ orbital is confined at the bottom of a quantum well and has the narrowest depth extension. The $d_{xz}/d_{yz}$ orbitals at higher energy will be less confined, and this is even more so for the higher energy $d_{x^2-y^2}$ and $d_{z^2}$ orbitals. In Figure 6(a), we show the experimental RCs (open circles) and the best-fit results (curves) of the RIXS excitations. We anticipate that in the SW-RIXS spectra, the observed $dd$ excitations should come from Ti 3$d$ interface carriers with associated structural defects and the fluorescence feature comes from the true 2DES. For these RCs, the intensity modulations of the RIXS excitations around the Bragg angle (~ 20°) are evidently different, which do not fit into the simulated RCs from the whole STO layer. The discrepancy means that these RIXS excitations have different spatial distributions.

In general, a RC with a larger intensity modulation corresponds to a more localized depth distribution. With this in mind, the intensity modulations in Figure 6(a) imply that the fluorescence feature has a wider depth distribution than the $dd$ excitations. It is intriguing to see different depth profiles for the $dd$ excitations, and to interpret such depth distribution, one can use the subband picture from the aforementioned quantum confinement effect.[63,66,68] From the analysis of RCs in Figure 6(a), we can obtain the depth distributions of these RIXS excitations which are summarized in Figure 6(b). The extent of the depth distributions of different RIXS excitations are as expected, and the SW-RIXS results give a consistent picture with SW-RXPS that there is a noticeably large amount of Ti 3$d$ interface carriers residing in both the LAO and STO sides of the interface. However, a more quantitative interpretation to include the RIXS cross section and the quantum confinement effects will be needed for the future SW-RIXS studies.

## V. Conclusion and future directions

In this review, we introduce the basic concepts of x-ray SW spectroscopies and discuss the design of SL to enhance the reflectivity and the SW effects. We review the development of SW-XPS and recent applications of SW-XPS/SW-RIXS on studying the interfacial phenomena in the polar-nonpolar perovskite oxide heterostructures. For the LCO/STO SLs, the built-in potential gradients across the interfaces caused by the polarity discontinuities can be fully visualized using resonant SW-XPS. For the LAO/STO SLs, SW-RXPS results suggest that two different types of Ti $3d$ interface carriers exist across the LAO/STO interfaces. One is the widely-discussed 2DES in the STO layer and the other is the novel polarity-induced Ti-on-Al antisite defects in the LAO layer. These findings provide a microscopic picture of how the localized and mobile Ti $3d$ interface carriers distribute across the LAO/STO interfaces. One future direction of SW-RXPS studies on the LAO/STO system is to distinguish the depth profiles of the IG and QP states. The QP and IG states are known to be associated with 2DES and localized electrons, respectively.[58,59] Discriminating these two states in both energy and real space regimes can help understand and control of the formation of 2DES. In addition, it will also shed light on the relationship between the oxygen vacancies and 2DES.

From the methodological perspective, SW-RIXS has demonstrated its potential in studying the interfacial distributions of RIXS excitations, such as the fluorescence, magnetic, and $dd$ excitations. Recent works that reported high quality epitaxial nickelate thin films[24,69,70] by MBE and PLD have drawn much attention. For example, the SrNiO$_3$/LaFeO$_3$ SL offer the evidence of a superior control of valence states from Ni$^{3+}$ to Ni$^{4+}$ through the thickness of constituent layers.[24] We think that this SL sample will be an ideal platform for SW-RIXS to visualize the depth profiles of Ni$^{3+}$ and Ni$^{4+}$ and study the length scales of the interface charge transfers.

In addition to the resonance effect, there are many parameters that one can use to control the interplay between the SW electric field and the photon-material interactions. A comprehensive study using a variety of x-ray SW techniques (e.g. SW-XPS, SW-RXPS, and SW-RIXS) with improved depth and interface sensitivity on model SLs is highly desirable. Such studies can provide more complete pictures of the interface electronic structure and pave ways for the functionality control of the perovskite oxide heterostructures. With rich emergent properties found in the TMO polar-nonpolar interfaces and the unique advantage of SW spectroscopies, we expect that SW spectroscopies will be a flourishing research field in the near future.

## ACKNOWLEDGMENTS

Use of the Stanford Synchrotron Radiation Lightsource, SLAC National Accelerator Laboratory, is supported by the U.S. Department of Energy, Office of Science, Office of Basic Energy Sciences under Contract No. DE-AC02-76SF00515. This research uses the resource of the Advanced Light Source, a U.S. DOE Office of Science User Facility under contract number DE-AC02-05CH11231.

## DATA AVAILABILITY

Data sharing is not applicable to this article as no new data were created or analyzed in this study.

## Figure captions

FIG.1. (a) Schematic of an $(LSCO_2/LSMO_7)_{20}$ superlattice sample for x-ray SW spectroscopic experiments, and (b) its corresponding SW electric field intensity map as a function of depth and incidence angle. (b) is reprinted with permission from C.-T. Kuo et al., Phys. Rev. B 98, 235146 (2018). Copyright 2018 American Physical Society.

FIG.2. (a) Calculated reflectivity at hν = 900 eV for LCO/STO SLs with different sample geometries. (b) Reflectivity of an $(LCO_5/STO_{10})_{10}$ SL at 700, 900, and 1100 eV incident photon energies.

FIG.3. (a) On-resonant (on-res) and off-resonant (off-res) reflectivity of an $(LSCO_2/LSMO_7)_{10}$ SL near the Cu $L_3$ XAS edge (~932 eV). (b) Resonant index of refraction ($β_{res}$ and $δ_{res}$) of an $(LSCO_2/LSMO_7)_{10}$ SL derived from the Kramers-Kronig analysis. The non-resonant index of refraction ($β_{non-res}$ and $δ_{non-res}$) from CXRO website are plotted for comparison.

FIG.4. (a) La $M_5$ x-ray absorption spectrum of an $(LCO_5/STO_{10})_{10}$ SL. (b) The real (δ) and the imaginary (β) parts of the index of refraction derived from the Kramers-Kronig analysis. The electric field strength distribution as a function of sample depth and incidence angle at (c) 829.7 eV and (d) 831.5 eV. Experimental and simulated relative energy shifts of Sr 3*d* and La 4*d* core levels versus incident angle at (e) 829.7 eV and (f) 831.5 eV. Experimental valence-band decompositions and the corresponding simulations at (g) 829.7

eV and (h) 831.5 eV. (i) SW-XPS-derived (turquoise curves) and DFT-calculated (black curves) depth-resolved valence-band maximum for the top three layers of the (LCO$_5$/STO$_{10}$)$_{10}$ SL. Reprinted with permission from S.-C. Lin et al., Phys. Rev. B 98, 165124 (2018). Copyright 2018 American Physical Society.

FIG.5. (a) An RXPS valence band spectrum at hν = 459 eV. The inset shows the magnified view around the Fermi level where a fitted NF peak can be seen at ~ 0.35 eV. (b) Experimental and theoretical rocking curves of La 4$d$ core level, Sr 3$d$ core levels, and the NF peak. (c) Illustration of the locations of ions that contribute to the NF peak. The lower panel is the determined depth profile which shows two main sources for the Ti 3$d$ interface states. One is the 2DES in the STO layer and the other is the Ti-Al antisite (Ti$_{Al}$) in the LAO layer. Reprinted with permission from S.-C. Lin et al., Phys. Rev. Materials 4, 115002 (2020). Copyright 2020 American Physical Society.

FIG.6. (a) Experimental (open circle) and best-fit (curve) rocking curves for $d_{xz}/d_{yz}$, fluorescence, $d_{x^2-y^2}$, and $d_{z^2}$ excitations in the SW-RIXS spectra. (b) Determined depth profiles for the RIXS excitations in the top five interfaces of an LAO/STO heterostructure. Reprinted with permission from S.-C. Lin et al., Phys. Rev. Materials 4, 115002 (2020). Copyright 2020 American Physical Society.

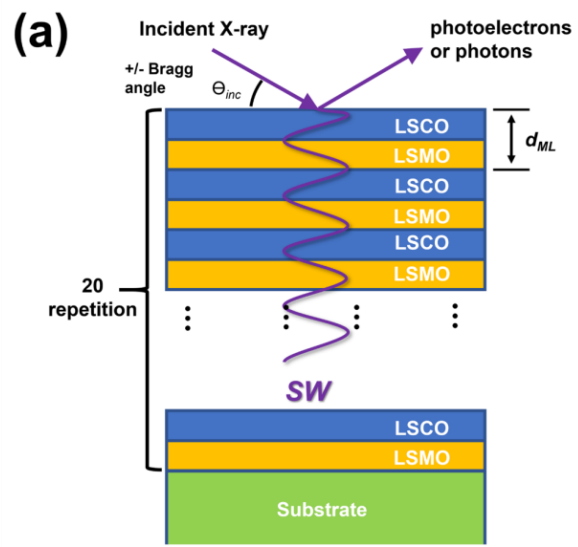

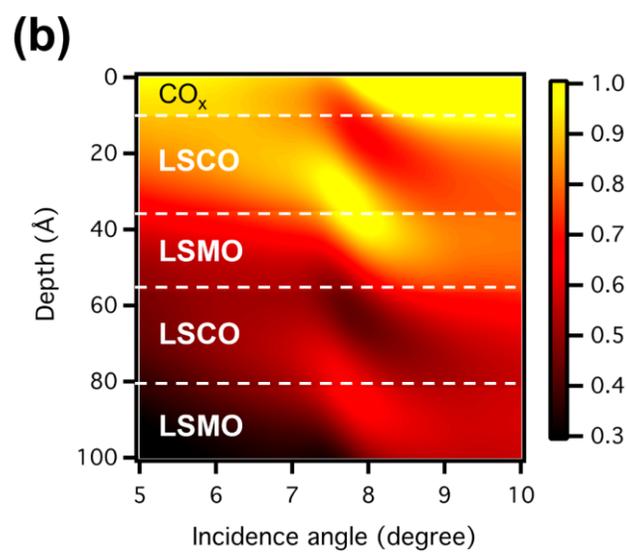

FIG. 1.

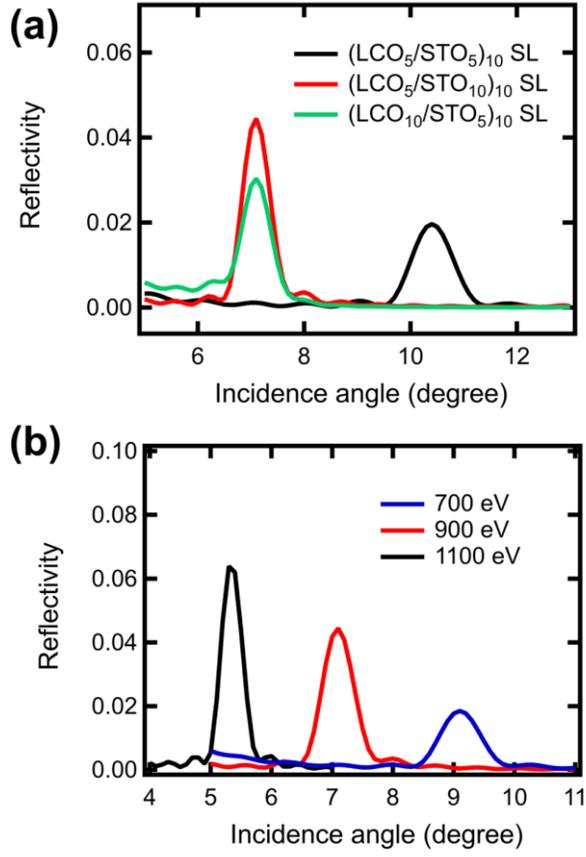

FIG. 2.

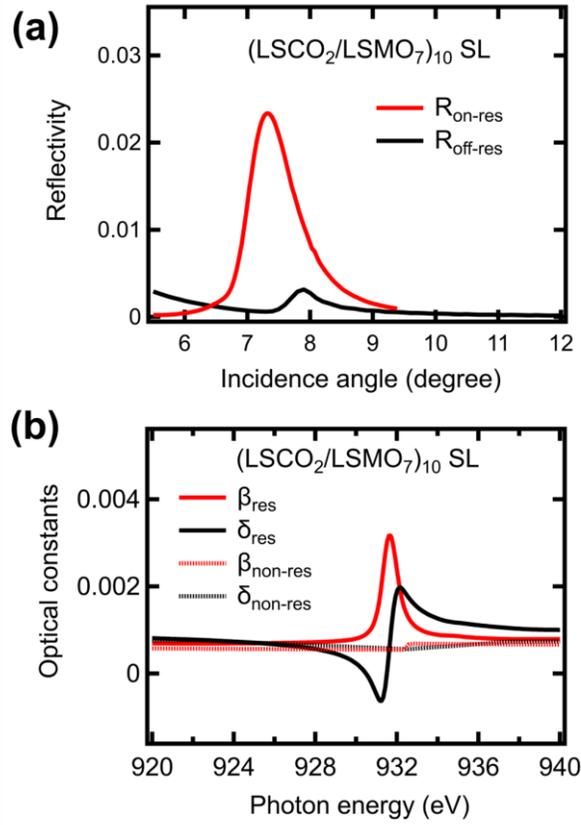

FIG. 3.

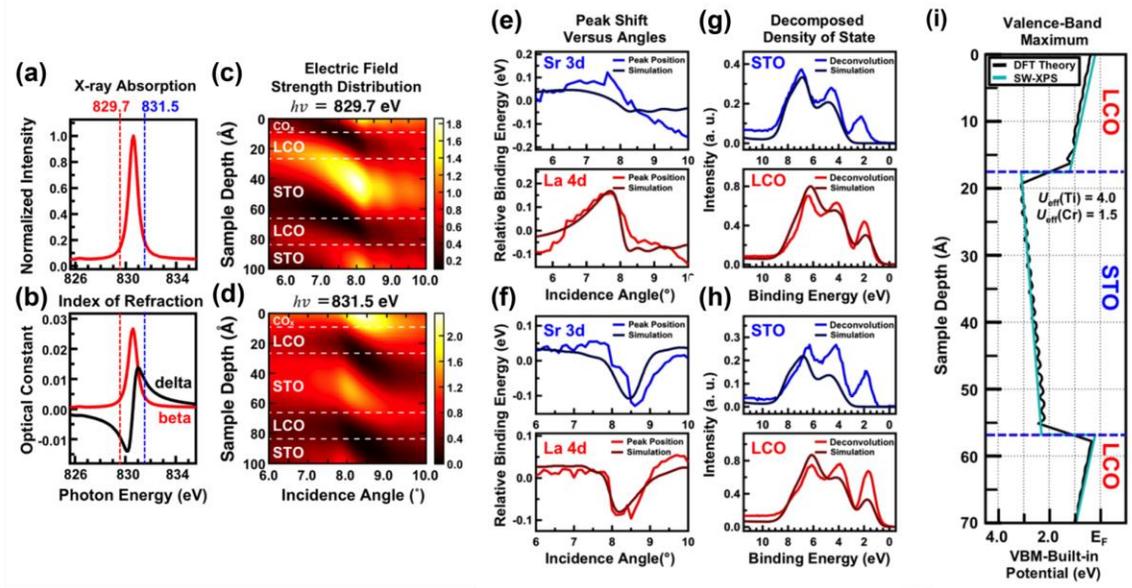

FIG. 4.

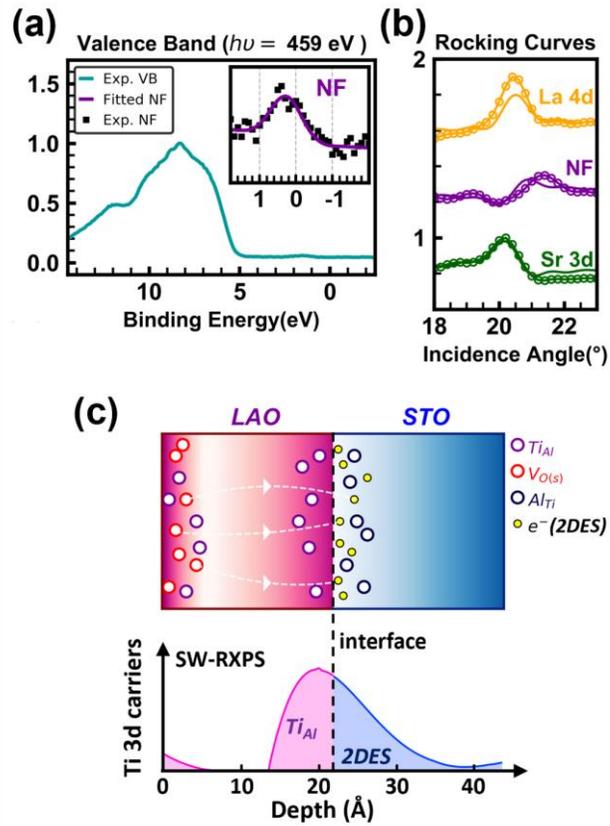

FIG. 5.

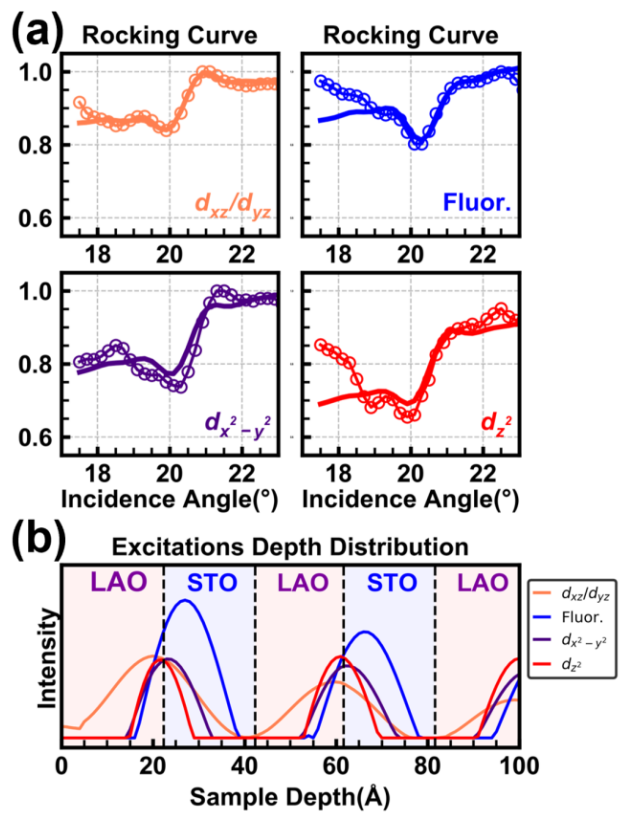

FIG. 6.